Preprint SMU-HEP-19-01

# Direct ellipsoidal fitting of discrete multi-dimensional data

Rafey Anwar, Madeline Hamilton, Pavel M. Nadolsky

Department of Physics, Southern Methodist University, Dallas, TX 75275-0175

May 15, 2019

## ABSTRACT

Multi-dimensional distributions of discrete data that resemble ellipsoids arise in numerous areas of science, statistics, and computational geometry. We describe a complete algebraic algorithm to determine the quadratic form specifying the equation of ellipsoid for the boundary of such multi-dimensional discrete distribution. In this approach, the equation of an ellipsoid is reconstructed using a set of matrix equations from low-dimensional projections of the input data. We provide a Mathematica program realizing the full implementation of the ellipsoid reconstruction algorithm in an arbitrary number of dimensions. To demonstrate its many potential uses, the direct reconstruction method is applied to quasi-Gaussian statistical distributions arising in elementary particle production at the Large Hadron Collider.

## I. Introduction

In this article, we focus on a common problem, reconstruction of a d-dimensional ellipsoid from coordinates of a set of discrete data points populating the volume of the ellipsoid. Clusters of data points that are approximately ellipsoidal in shape are encountered in many applications ranging from multivariate statistical analysis and machine learning to cardiac strain imaging (1) and calibration of magnetic compasses (2). Given the images (projections) of the ellipsoid, the task is to find the equation of the ellipsoid's surface in a suitable coordinate representation. Remarkably, the equation of such an ellipsoid can be found by analytically solving a system of matrix equations, as described below.

For example, suppose N discrete predictions dependent on parameters $\{x_1, x_2, \ldots, x_d\}$ are distributed in an approximately ellipsoidal region in the d-dimensional parameter space. In statistical analysis, these predictions can be generated by random sampling from a multi-dimensional probability distribution that is approximately Gaussian. If the equation specifying the underlying probability distribution is unknown, one might wish to reconstruct it from the discrete distribution of the data. One way of doing this is to select points on the boundary of the d-dimensional region satisfying a given probability level and fit an ellipsoid to this boundary. From the quadratic form describing the ellipsoid, the quasi-Gaussian probability distribution can be immediately determined.

A practical algorithm for the reconstruction of a d-dimensional ellipsoid by fitting discrete points was developed by Bertoni (3). It is based on the combination of methods developed by Fitzgibbon, Pilu, Fisher (4) and Karl (5). Bertoni's algorithm is general, allowing one to reconstruct an ellipsoid from a complete set of the low-dimensional (not necessarily independent) ellipsoid's projections. However, Karl's and Bertoni's papers do not demonstrate existence of a unique solution. In fact, such solution exists only when the set of projections is sufficiently complete to determine all coefficients of the quadratic form.

In this article, we focus on a special case, when the ellipsoid is reconstructed from its two-dimensional orthogonal projections. We show how to derive a closed solution for the ellipsoid's quadratic form using a set of complete and mutually consistent two-dimensional projections. The existence of such a unique solution, and the algebraic formula to find its coefficients, is a new result presented below. A Mathematica program implementing the full reconstruction algorithm is available upon request.

The reconstruction algorithm has important applications in the field of elementary particle physics. For example, the structure of protons and nuclei in high-energy collisions is parameterized by parton distribution functions (PDFs) that are determined from a large-scale multivariate analysis of experimental measurements (6). To determine theoretical uncertainties for the rates of elementary particle production at the Large Hadron Collider, one may need to reconstruct an underlying quasi-Gaussian probability distribution from the multidimensional distribution of values obtained by stochastic sampling. Traditionally, the Gaussian distribution can be estimated using the method of the covariance matrix (7) or related Hessian matrix (8). Our ellipsoid reconstruction algorithm can be employed as a part of an alternative estimation method that does not assume that the probability distribution is perfectly Gaussian, as we explain in Section 4. To demonstrate the usefulness of the developed reconstruction method and explore its differences against the covariance matrix method, we employ both methods to predict the uncertainty due to PDF parameterizations in production of $W^{\pm}$, $Z^{0}$, and $H^{0}$ bosons at the LHC.

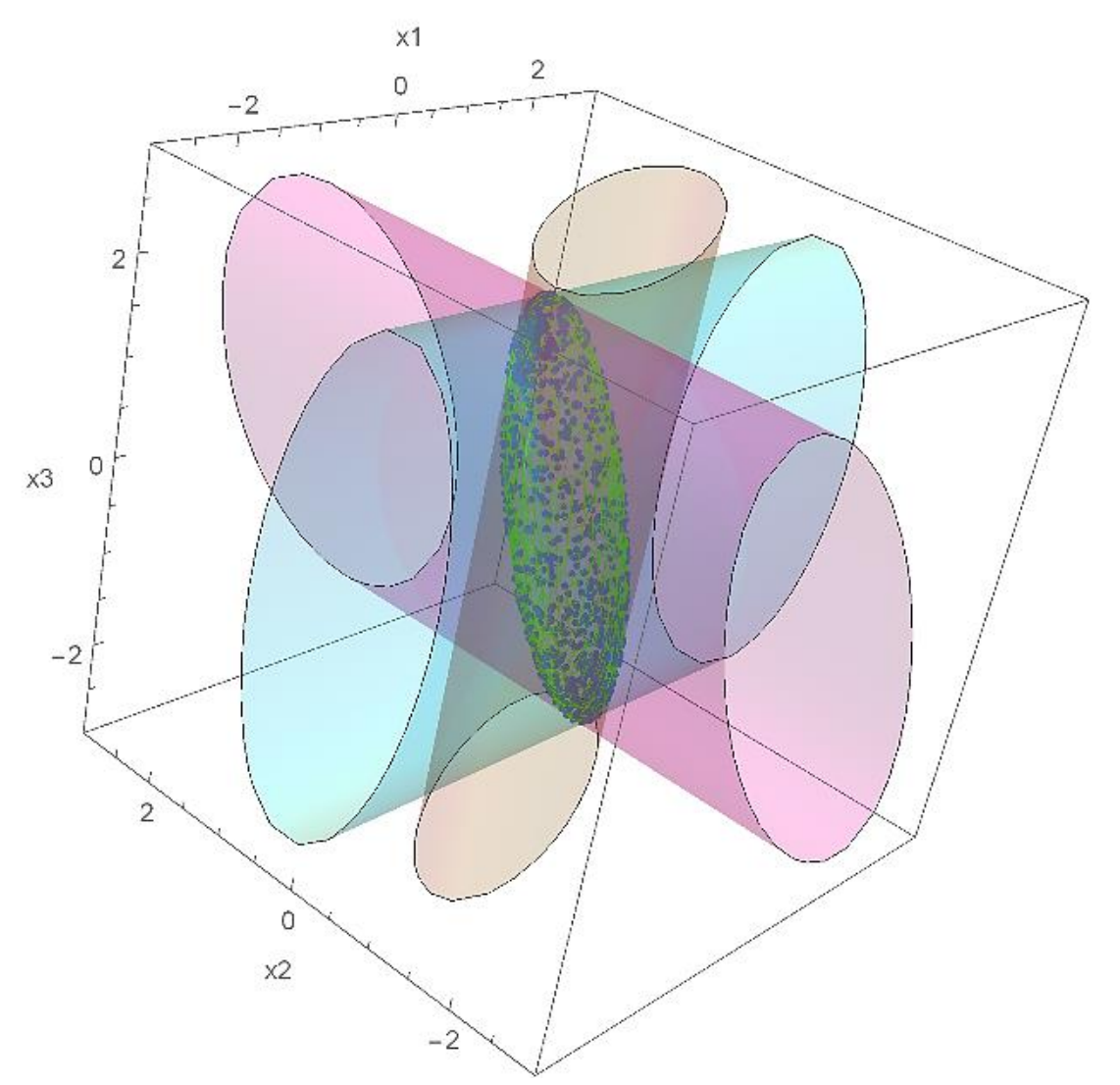

*Figure 1. A three-dimensional ellipsoid fitted to 1000 quasi-ellipsoidal points and its two-dimensional elliptical projections.*

Figure 1 illustrates the reconstruction of a 3-dimensional ellipsoid from its 2-dimensional elliptical projections. The input data consists of 1000 random three-dimensional vectors (blue points) that populate the ellipsoid's volume. The output consists of the *3x3* symmetric matrix $A_3$ specifying the equation of the ellipsoid boundary (shown by a green mesh), found from the discrete input data with the help of our method. The first step is to project the input vectors onto independent orthogonal planes, where the boundaries of the input clusters are fitted by ellipses, as described in Section II. Then, in Section III, we reconstruct the output matrix $A_3$ from the matrices $A_{2,i}$ ($i = 1,2,3$) for the equations of the projected ellipses. This Section presents a general formula for reconstructing the d-dimensional ellipsoid matrix $A_d$ from the 2-dimensional projection matrices $A_{2,i}$, where $1 \leq i \leq d(d-1)/2$. It also provides a proof that such a matrix exists and a consistency check for the projection matrices. Section IV applies the Mathematica program to the analysis of production cross sections in elementary particle physics. Section V contains our conclusions.

## II. Fitting two-dimensional ellipses

As the first step in the reconstruction of the d-dimensional ellipsoid, we need to determine the matrices for the boundaries of two-dimensional ellipses that are the projections of the ellipsoid onto the orthogonal two-dimensional planes. In the example of Figure 1, the projected input data vectors populate the inside of an ellipse in each projection plane. The Convex Hull (CH) Method described in Subsection A reconstructs the quadratic form for the convex boundary of this ellipse with the help of the least-squares elliptical fitting algorithm described in (4).

In statistical applications, the cluster of data vectors sampled from a quasi-Gaussian distribution does not have a sharp boundary. Rather, the “ellipse” may correspond to the boundary of the probability-$\alpha$ region determined from the covariance matrix (CM) according to the conventional method summarized in Subsection B.

## A. The Convex Hull Method

Since a 2-dimensional projection of a d-dimensional ellipsoid represents a filled ellipse and not its outline, one must first find the boundary, or convex hull, of the projection and then fit an ellipse to this boundary. The convex hull algorithm addresses the first necessity, while the least squares elliptical fitting algorithm addresses the second.

### 1. Finding the convex hull

The convex hull algorithm determines which points of the data set would be most appropriate for use in elliptical fitting, so that the resulting ellipse describes the boundary of the data subset, not the data subset as a whole. We will describe a convex hull algorithm that operates with cross products, although other algorithms for convex hull reconstruction are also available, such as the one described in (9).

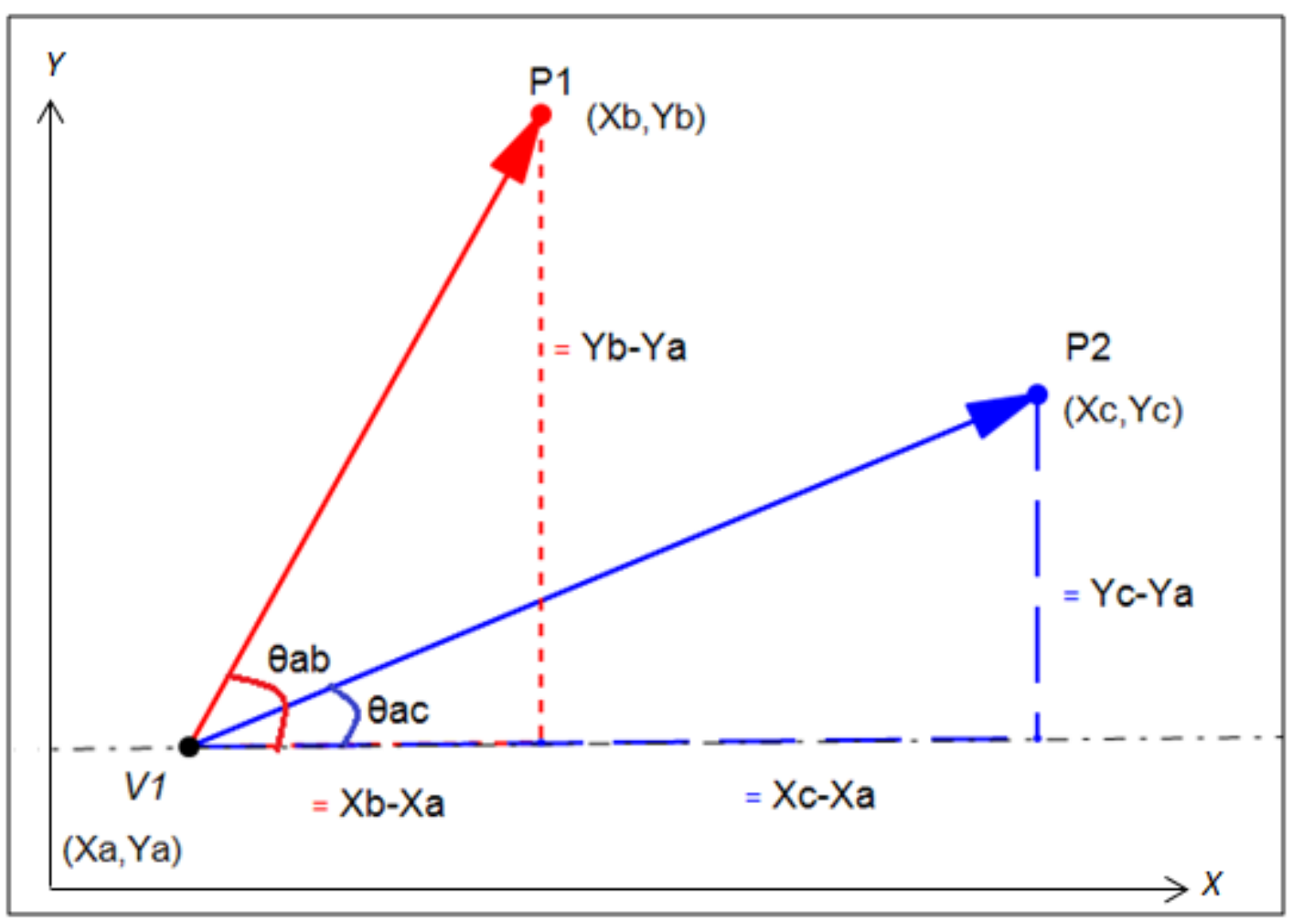

*Figure 2: Illustration of the vectors arising in the determination of the convex hull*

A convex hull of a set of points in an *xy* plane is the smallest convex polygon in the plane that contains every point in the set. For visualization purposes, it can be described as the shape that a rubber band would take if it were stretched out around a set of points. The first step in finding the convex hull of a set of points is finding the convex hull's vertices. These vertices are points from the data set such that if they were connected by straight lines, the polygon formed would be the convex hull of the data set.

To begin, a point $V_1$ from the set known to be a vertex is needed. If such a point is not explicitly given, it can easily be found by taking the point with the lowest $x$ value, as this point will certainly be a vertex due to its extreme position. $V_1$ is now the active vertex. To find the next vertex, the active vertex is used as a basis of comparison for every other point in the set. Whichever point in the set creates the greatest angle relative to $V_1$'s horizontal axis over $[0, \pi]$ will be the next vertex, $V_2$. $V_2$ will now act as the active vertex to find $V_3$. This will continue until a point $V_n$ whose next vertex is $V_1$ , the original active vertex, is found. Once this point has been reached, all the vertices $\{V_1, V_2, \ldots, V_n\}$ of the convex hull have been found. Connecting these vertices with straight lines creates the convex hull.

The algorithm can be explained in detail using Figure 2. In the figure, *V1* is the active vertex, and *P1* and *P2* represent the two points currently being compared. $\theta_{ab}$ and $\theta_{ac}$, the angles that are compared for each pair of points, can be calculated as follows:

$$\Theta ab = \tan^{-1}\left(\frac{Yb - Ya}{Xb - Xa}\right), \qquad \Theta ac = \tan^{-1}\left(\frac{Yc - Ya}{Xc - Xa}\right).$$

However, because trigonometric functions are computationally slow, simpler algebraic representations of the angles are used, and the following test is obtained:

$$(Yb - Ya)(Xc - Xa) - (Yc - Ya)(Xb - Xa) = \delta.$$

This test returns a determinant δ. *Xa* and *Ya* represent the coordinates of the current active vertex, $V_A$. *Xb* and *Yb* represent the coordinates of any point $P_B$ , and *Xc* and *Yc* represent the coordinates of any point $P_C$. If $\delta > 0$, then $P_B$ creates the larger angle with respect to $V_A$. If $\delta < 0$, then $P_C$ creates the larger angle with respect to $V_A$. If the determinant is zero, then all three points are collinear and the point which is farther from $V_A$ should be selected.

It is also easy to realize that δ represents the *z* component of the cross product $\overrightarrow{AB} \times \overrightarrow{AC}$, so that $\delta > 0$ ($\delta < 0$) represents the clockwise (counterclockwise) rotation of $\overrightarrow{AB}$ toward $\overrightarrow{AC}$, which can also be used to determine the relative orientation of $\overrightarrow{AB}$ and $\overrightarrow{AC}$. The program repeats this process as needed until all the convex hull values have been found.

## 2. Fitting an ellipse to the convex hull

Next, we need to find an ellipse that would provide a reasonable fit to the points on the convex hull.

If a point lies on an ellipse, the point's coordinates satisfy

$$F(\boldsymbol{A}, X) = a_1x^2 + a_2xy + a_3y^2 + a_4x + a_5y + a_6 = 0,$$

where the coefficients are constrained by $a_2{}^2 - 4a_1a_3 < 0$. For $n$ points $\{x_1, y_1\}, \dots, \{x_n, y_n\}$ that are not exactly on the ellipse, the desired ellipse can be obtained through a least squares minimization of algebraic distances from the points to the ellipse. As explained in (4), the minimization problem for finding the ellipse can be expressed as a generalized eigenvalue problem based on a matrix equation

$$\boldsymbol{SA} = \lambda\boldsymbol{CA} \qquad \text{(Eq. 1)}$$

where $\lambda$ is the eigenvalue, $\boldsymbol{A} = (a_1\ a_2\ a_3\ a_4\ a_5\ a_6)^T$, and $\boldsymbol{S}$ and $\boldsymbol{C}$ are certain $6 \times 6$ matrices constructed in Ref. (4).

The generalized eigenvalue problem can be solved numerically using LAPACK (10), Mathematica, or another advanced linear algebra package. Alternatively, it is possible to reduce this equation to a standard eigenvalue problem using the method that will be now described. This method can be easily implemented with any linear algebra library.

Toward this goal, we identify two 3-component vectors

$$\vec{a}_1 = \begin{pmatrix} a_1 \\ a_2 \\ a_3 \end{pmatrix} \quad \text{and} \quad \vec{a}_2 = \begin{pmatrix} a_4 \\ a_5 \\ a_6 \end{pmatrix}$$

containing the coefficients associated with rotations and translations inside the 6-component vector $\boldsymbol{A} = [\vec{a}_1 \quad \vec{a}_2]^T$. Block matrices are indicated by bold letters and square brackets. Express matrices $\boldsymbol{S}$ and $\boldsymbol{C}$ in terms of $3 \times k$ and $k \times 3$ blocks, where $k = 1$ or $3$:

$$\boldsymbol{S} = \begin{bmatrix} S_{11} & S_{12} \\ S_{21} & S_{22} \end{bmatrix}, \quad \boldsymbol{C} = \begin{bmatrix} M & 0_3 \\ 0_3 & 0_3 \end{bmatrix},$$

with

$$M = \begin{pmatrix} 0 & 0 & 2 \\ 0 & -1 & 0 \\ 2 & 0 & 0 \end{pmatrix},$$

and a $3 \times 3$ zero matrix $0_3$.

If $\boldsymbol{D} = [d_1 \quad d_2]$ with

$$d_1 = \begin{pmatrix} x_1^2 & x_1y_1 & y_1^2 \\ x_2^2 & x_2y_2 & y_2^2 \\ \cdots & \cdots & \cdots \\ x_n^2 & x_ny_n & y_n^2 \end{pmatrix} \quad \text{and} \quad d_2 = \begin{pmatrix} x_1 & y_1 & 1 \\ x_2 & y_2 & 1 \\ \cdots & \cdots & \cdots \\ x_n & y_n & 1 \end{pmatrix},$$

the $3 \times 3$ blocks $S_{ij}$ of $\boldsymbol{S}$ are given according to Ref. (4) by $S_{ij} = d_i^T d_j$. Eq. 1 can then be written as

$$\begin{bmatrix} S_{11} & S_{12} \\ S_{21} & S_{22} \end{bmatrix} \begin{pmatrix} \vec{a}_1 \\ \vec{a}_2 \end{pmatrix} = \lambda \begin{bmatrix} M & 0_3 \\ 0_3 & 0_3 \end{bmatrix} \begin{pmatrix} \vec{a}_1 \\ \vec{a}_2 \end{pmatrix}. \qquad \text{(Eq. 2)}$$

We apply singular value decomposition to $\boldsymbol{C}$ to find

$$\boldsymbol{C} = \boldsymbol{U L_I L_0 V^T}, \qquad \text{(Eq. 3)}$$

which depends on block matrices

$$\boldsymbol{U} = \begin{bmatrix} u_3 & 0_3 \\ 0_3 & I_3 \end{bmatrix}, \qquad \boldsymbol{V} = \begin{bmatrix} v_3 & 0_3 \\ 0_3 & I_3 \end{bmatrix}, \qquad \boldsymbol{L_I} = \begin{bmatrix} l_3 & 0_3 \\ 0_3 & I_3 \end{bmatrix}, \qquad \boldsymbol{L_0} = \begin{bmatrix} I_3 & 0_3 \\ 0_3 & 0_3 \end{bmatrix}.$$

In this equation, $I_3$ is the $3 \times 3$ identity matrix,

$$u_3 = \begin{pmatrix} 1 & 0 & 0 \\ 0 & 0 & -1 \\ 0 & 1 & 0 \end{pmatrix}, \qquad v_3 = \begin{pmatrix} 0 & 1 & 0 \\ 0 & 0 & 1 \\ 1 & 0 & 0 \end{pmatrix}, \qquad l_3 = \begin{pmatrix} 2 & 0 & 0 \\ 0 & 2 & 0 \\ 0 & 0 & 1 \end{pmatrix}. \qquad \text{(Eq. 4)}$$

The only singular matrix in Eq. 3 is $\boldsymbol{L_0}$: $\det \boldsymbol{L_0} = 0$. On the other hand, $\boldsymbol{U}$ and $\boldsymbol{V}$ are orthonormal, $\boldsymbol{UU^T} = \boldsymbol{VV^T} = I_6$. The inverse of $\boldsymbol{L_I}$ also exists,

$$\boldsymbol{L_I^{-1}} = \begin{bmatrix} l_3^{-1} & 0_3 \\ 0_3 & I_3 \end{bmatrix}, \text{ where } l_3^{-1} = \begin{pmatrix} 1/2 & 0 & 0 \\ 0 & 1/2 & 0 \\ 0 & 0 & 1 \end{pmatrix}.$$

In this representation, the only complication is associated with the singular $\boldsymbol{L_0}$ matrix inside the decomposition for $\boldsymbol{C}$. We therefore multiply Eq. 1 by $\boldsymbol{U}^T \boldsymbol{L}_I^{-1}$ from the left and identify $\boldsymbol{S}_V \equiv \boldsymbol{U}^T \boldsymbol{L}_I^{-1} \boldsymbol{S}\, \boldsymbol{V}$, $\vec{\boldsymbol{A}}_V \equiv \boldsymbol{V^T} \vec{\boldsymbol{A}}$ to obtain

$$\boldsymbol{S}_V \vec{\boldsymbol{A}}_V = \lambda \boldsymbol{L_0} \vec{\boldsymbol{A}}_{\boldsymbol{V}}.$$

In the block form, this equation is

$$\begin{bmatrix} S_{V11} & S_{V12} \\ S_{V21} & S_{V22} \end{bmatrix} \begin{pmatrix} \vec{a}_{V1} \\ \vec{a}_{V2} \end{pmatrix} = \begin{bmatrix} \lambda & 0_3 \\ 0_3 & 0_3 \end{bmatrix} \begin{pmatrix} \vec{a}_{V1} \\ \vec{a}_{V2} \end{pmatrix},$$

and

$$\begin{bmatrix} S_{V11} & S_{V12} \\ S_{V21} & S_{V22} \end{bmatrix} = \begin{bmatrix} l_3^{-1} u_3^T S_{11} v_3 & l_3^{-1} u_3^T S_{12} \\ S_{21} v_3 & S_{22} \end{bmatrix}.$$

To proceed, we need to single out a special case when all points lie on a single line, corresponding to a degenerate solution for the elliptical coefficients. It can be easily demonstrated that the points lie on a line if and only if $\det S_{22} = 0$. Indeed, since $S_{22} = d_2^T d_2$, the condition $\det S_{22} = 0$ is equivalent to $\det d_2 = 0$. Then, there is a vector $w = (w_x \quad w_y \quad 1)^T$ such that

$$d_2 w = \begin{pmatrix} x_1 & y_1 & 1 \\ x_2 & y_2 & 1 \\ \cdots & \cdots & \cdots \\ x_n & y_n & 1 \end{pmatrix} \begin{pmatrix} w_x \\ w_y \\ 1 \end{pmatrix} = \begin{pmatrix} w_x x_1 + w_y y_1 + 1 \\ w_x x_2 + w_y y_2 + 1 \\ \cdots \\ w_x x_n + w_y y_n + 1 \end{pmatrix} = \begin{pmatrix} 0 \\ 0 \\ \cdots \\ 0 \end{pmatrix};$$

or, all points lie on the line $w_x x + w_y y + 1 = 0$.

If the solution is not degenerate ($\det S_{22} \neq 0$), the system of equations becomes

$$(S_{V11} - S_{V12} S_{V22}^{-1} S_{V21}) \vec{a}_{V1} = \lambda \vec{a}_{V1}, \qquad \text{(Eq. 5)}$$

$$\vec{a}_{V22} = -S_{V22}^{-1} S_{21} \vec{a}_{V1}. \qquad \text{(Eq. 6)}$$

The first equation is a regular eigenvalue problem for $\vec{a}_{V1}$, solved by standard methods. The second equation derives $\vec{a}_{V22}$ from $\vec{a}_{V1}$.

For a degenerate solution ($\det S_{22} = 0$), it suffices to fit all points using linear regression (assuming that $\vec{a}_{V1}$ is a null vector).

Based on this exposition, the equation of the ellipse is found as follows. Given the coordinates $\{x_k, y_k\}$ of the fitted points, we compute the matrices $d_i$, $S_{ij}$, and $S_{Vij}$ for $i, j = 1$ or 2. The determinant $\det S_{22}$ is calculated to decide if the points lie on one line within the accuracy of the calculation. If the solution is not degenerate, we find $\vec{a}_{V1}$ and $\vec{a}_{V2}$ from Eqs. 5 and 6; otherwise, we set $a_{Vi} = 0$ for $i = 1,2,3$ and find $a_{V4}$ and $a_{V5}$ by linear regression.

Among three possible eigenvalues in Eq. 5, one eigenvalue is positive and two are negative (4). The eigenvector $\vec{a}_{V1}$ that solves our conic problem corresponds to the only positive eigenvalue (4).[1]

Finally, the coefficient vectors are determined as

$$\vec{a}_1 = N\, v_3 \vec{a}_{V1}, \quad \vec{a}_2 = N\, \vec{a}_{V2},$$

where $N$ is an arbitrary normalization factor that can multiply all coefficients $a_i$ without violating the original equation $F(\boldsymbol{A}, X) = 0$. $N$ must be found from a supplementary condition. For example, it can be that the quadratic form for the ellipse has a standard normalization so that at the center of the ellipse, $X_0 = \{x_0, y_0\}$, the quadratic form takes the value $F_{standard}(\boldsymbol{A}, X_0) = -1$. The coordinates of the center can be found from $a_i$ as

$$x_0 = -\frac{a_2 a_5 - a_3 a_4}{a_2^2 - 4 a_1 a_3}, \quad y_0 = -\frac{a_2 a_4 - a_1 a_5}{a_2^2 - 4 a_1 a_3}$$

independently of $N$. Then, once $\{x_0, y_0\}$ is determined using $N = 1$ for $a_i$, the final normalization that satisfies $F(\boldsymbol{A}, X_0, N_{final}) = -1$ is obtained by

$$N_{final} = -\frac{1}{F(\boldsymbol{A}, X_0, N = 1)}.$$

If the ellipse is centered at the origin, $x_0 = y_0 = 0$, the final equation of the ellipse in the convex hull (CH) method is

[1] If there is not enough noise in the data (all points lie exactly on the ellipse), the positive eigenvalue may be indistinguishable from zero within accuracy. In this case, the solution corresponds to the largest eigenvalue.

$$(x \quad y) \cdot A_2^{CH} \cdot \begin{pmatrix} x \\ y \end{pmatrix} = 1, \text{ with } A_2^{CH} \equiv -\begin{pmatrix} a_1 & \frac{a_2}{2} \\ \frac{a_2}{2} & a_3 \end{pmatrix}. \qquad \text{Eq. (7)}$$

### B. The Covariance Matrix Method

If a sufficiently large sample of the two-dimensional data $\{(x_{1i} = x_i,\ x_{2i} = y_i)\}$ is drawn from a Gaussian distribution, another method, which we will refer to as the "Covariance Matrix" (CM) method, can be used to determine the ellipse that delineates the boundary of the region containing the fraction $\alpha$ of the data sample ($0 \leq \alpha \leq 1$). In the absence of any correlation, and under a simplifying assumption that the data have zero mean values, $\langle x_{1,2} \rangle = 0$, the points on the axis-aligned boundary ellipse would adhere to

$$\frac{x_1^2}{{\sigma_1}^2} + \frac{{x_2}^2}{{\sigma_2}^2} = s,$$

where $\sigma_{1,2} \equiv \sqrt{\langle x_{1,2}^2 \rangle}$ are the standard deviations of the $x_1$ and $x_2$ data, respectively, and $s$ is the chi-squared critical value associated with the desired probability level $\alpha$. On the other hand, if there is a correlation between $x_1$ and $x_2$, the resulting ellipse will no longer be aligned with the $x_1$ and $x_2$ axes and will satisfy

$$\frac{x_1^2}{{\sigma_1}^2} + \frac{x_2^2}{{\sigma_2}^2} - 2\,\frac{x_1}{\sigma_1}\frac{x_2}{\sigma_2}\cos\varphi = s\ \sin^2\ \varphi,$$

with

$$\cos\varphi \equiv \frac{\langle x_1 x_2 \rangle}{\langle x_1 \rangle \langle x_2 \rangle}.$$

This equation can be re-written using the same sign convention as in the previous subsection as

$$\sum_{i,j=1,2} x_i A_{2,ij}^{CM} x_j - s \sin^2 \varphi = 0$$

in terms of the matrix

$$A_2^{CM} \equiv \begin{pmatrix} \frac{1}{\sigma_1^2} & -\frac{\cos(\varphi)}{\sigma_1 \sigma_2} \\ -\frac{\cos(\varphi)}{\sigma_1 \sigma_2} & \frac{1}{\sigma_2^2} \end{pmatrix}. \qquad \text{Eq. (8)}$$

In contrast to the convex hull method, the matrix $A_2^{CM}$ in the CM method is shared by the entire input data, and the probability regions are distinguished only by the critical parameter $s$. This reflects the assumption behind the CM method that the probability distribution is exactly Gaussian. We also notice that the CM method implies that the correlation ellipsoid can be found directly by diagonalizing the covariance matrix in $d$ dimensions (without taking projections), i.e., by the $d$ −dimensional principal component analysis [PCA].

On the other hand, if we do not wish to use the CM method or suspect that the ellipsoid matrices may non-trivially depend on the probability level because of some deviations from the Gaussianity, the elliptical projection corresponding to the probability level $\alpha$ can be determined using the convex hull (CH) method, by first identifying a two-dimensional region containing a fraction $\alpha$ of the input data points, and then fitting an ellipse to the convex hull boundary of the enclosed data subset in this region. If the probability distribution deviates from the Gaussian one, the elliptical projections obtained with the CH method for different probability levels are not related by a simple rescaling of the parameter $s$. The comparison of the ellipsoids determined with the CH and CM methods thus provides a normality test for the underlying probability distribution.

## III. Reconstructing the ellipsoid from its projections

Next, we turn to the reconstruction of the ellipsoid from its two-dimensional projections. Notice that $d(d-1)/2$ independent projections are necessary to find all coefficients of the ellipsoid's quadratic form. The easiest way to proceed, then, is to determine, block-by-block, the *inverse* matrix of the quadratic form by repeatedly invoking Eq. 10 below for each projection. Here we lean on the crucial observation in Ref. (11) that it is the inverse matrices of the quadratic forms, rather than the quadratic forms themselves, that are straightforwardly related. Below we include a short proof of this important relation. [Ref. (8) presented a relation between the inverse matrices up to an overall normalization of their coefficients and without including a proof]. We bypass the difficulty of dealing with non-invertible operators that would affect, e.g., the direct implementation of the ellipsoid reconstruction method proposed by Karl (5). Karl's proposal requires stacking multiple projection operators in a way as to allow reconstruction of all ellipsoid's elements without omissions or double-counting. This is not necessary for the complete set of orthogonal projections, when the straightforward implementation using Eq. 10 is sufficient.

Any $d$-dimensional vector $\vec{x} = \{x_1, x_2, \dots, x_d\}$ drawn from the center of the ellipsoid to its surface satisfies

$$\vec{x}^T \cdot A_d \cdot \vec{x} = 1,$$

where $A_d$ is the matrix of the $d$-dimensional quadratic form whose elements we intend to find. A projection of $\vec{x}$ from $d$ to 2 dimensions, denoted by

$$\bar{\vec{x}} \equiv P_{2\leftarrow d} \cdot \vec{x}\,, \qquad \text{Eq. (9)}$$

obeys an analogous equation

$$\bar{\vec{x}}^T \cdot A_2 \cdot \bar{\vec{x}} = 1.$$

$A_2$ is the $2 \times 2$ matrix of the quadratic form for the projection found using the CH or CM method. $P_{2\leftarrow d}$ is a $2 \times d$ projection matrix, such as

$$P_{2\leftarrow d} = (\mathbb{I}_{2\times 2} \quad \mathbb{O}_{2\times d-2})$$

for the projection on the $x_1 x_2$ plane, with $\mathbb{I}_{2\times 2}$ and $\mathbb{O}_{2\times d-2}$ being the $2 \times 2$ identity matrix and $2 \times (d-2)$ zero matrix, respectively.

To put together $A_d$, we notice that the *inverse* matrices are related by

$$P_{2\leftarrow d} \cdot A_d^{-1} \cdot P_{2\leftarrow d}^T \;=\; A_2^{-1}. \qquad \text{Eq. (10)}$$

To prove it, recast the positive-definite symmetric matrix $A_d$ in terms of its eigenvalues $\lambda_i^2 > 0$ and the rotation matrix $O$,

$$A_d = O^T \Lambda^T \Lambda O \equiv \left(A_d^{1/2}\right)^T \cdot A_d^{1/2},$$

where $O^T O = \mathbb{I}_{d\times d}$, $\Lambda \equiv \mathrm{diag}(\lambda_1, \lambda_2, \dots, \lambda_d)$, and $A_d^{1/2} \equiv \Lambda O$.

$A_d^{1/2}$ generates an isomorphism that maps $\vec{x}$ onto a unit vector $\vec{n} = A_d^{1/2}\vec{x}$ satisfying $\vec{n}^T \cdot \vec{n} = 1$. In other words, the affine transformation specified by $A_d^{1/2}$ associates any $\vec{x}$ ending on the ellipsoid's surface to a vector $\vec{n}$ from the origin to a unit sphere. The inverse transformation also exists:

$$\vec{x} = A_d^{-\frac{1}{2}} \cdot \vec{n}. \qquad \text{Eq. (11)}$$

Similarly, the projections $\tilde{\vec{x}}$ are related to the projections $\tilde{\vec{n}} \equiv P_{2\leftarrow d} \cdot \vec{n}$ by

$$\tilde{\vec{x}} = A_2^{-\frac{1}{2}} \cdot \tilde{\vec{n}}. \qquad \text{Eq. (12)}$$

From Eqs. 9, 11, and 12, we conclude that

$$P_{2\leftarrow d} \cdot A_d^{-\frac{1}{2}} = A_2^{-\frac{1}{2}} \cdot P_{2\leftarrow d}.$$

Multiplying both sides by their transpose matrices on the right, and using $P_{2\leftarrow d} \cdot P_{2\leftarrow d}^T = \mathbb{I}_{2\times 2}$, we arrive at the desired relation,

$$P_{2\leftarrow d} \cdot A_d^{-1} \cdot P_{2\leftarrow d}^T \;=\; A_2^{-1}. \qquad \text{Q.E.D.}$$

In our practical algorithm, Eq. 10 is used to read off the coefficients of $A_d^{-1}$ directly from the coefficients of $A_2^{-1}$. If we generate $d(d-1)/2$ projections on planes $x_i x_j$ with $1 \le i \le d,\ i < j \le d$, the diagonal elements $(A_d^{-1})_{ii}$ will be equal to the diagonal elements in $(d-1)$ projections, and an off-diagonal element $(A_d^{-1})_{ij}$ will appear once in the projection $x_i x_j$ $(i \neq j)$. Due to noise, the $(d-1)$ computations of each diagonal element will not necessarily be exactly equivalent. The final estimate of a diagonal element is simply taken to be the mean value of the computations, and a comparison of the diagonal elements from the projections via their standard deviations and mean values provides a test of mutual consistency of the input projections.

A straightforward generalization of Eq. 10 relates the $d$-dimensional matrix $A_d$ to the matrices $A_{d'}$ of ellipsoids in lower-dimensional projections $(d' < d)$ using $d' \times d$ projection operators $P_{d'\leftarrow d}$:

$$P_{d'\leftarrow d} \cdot A_d^{-1} \cdot P_{d'\leftarrow d}^T \;=\; A_{d'}^{-1}. \qquad \text{Eq. (13)}$$

## IV. Applications

### A. A solid 3-dimensional ellipsoid

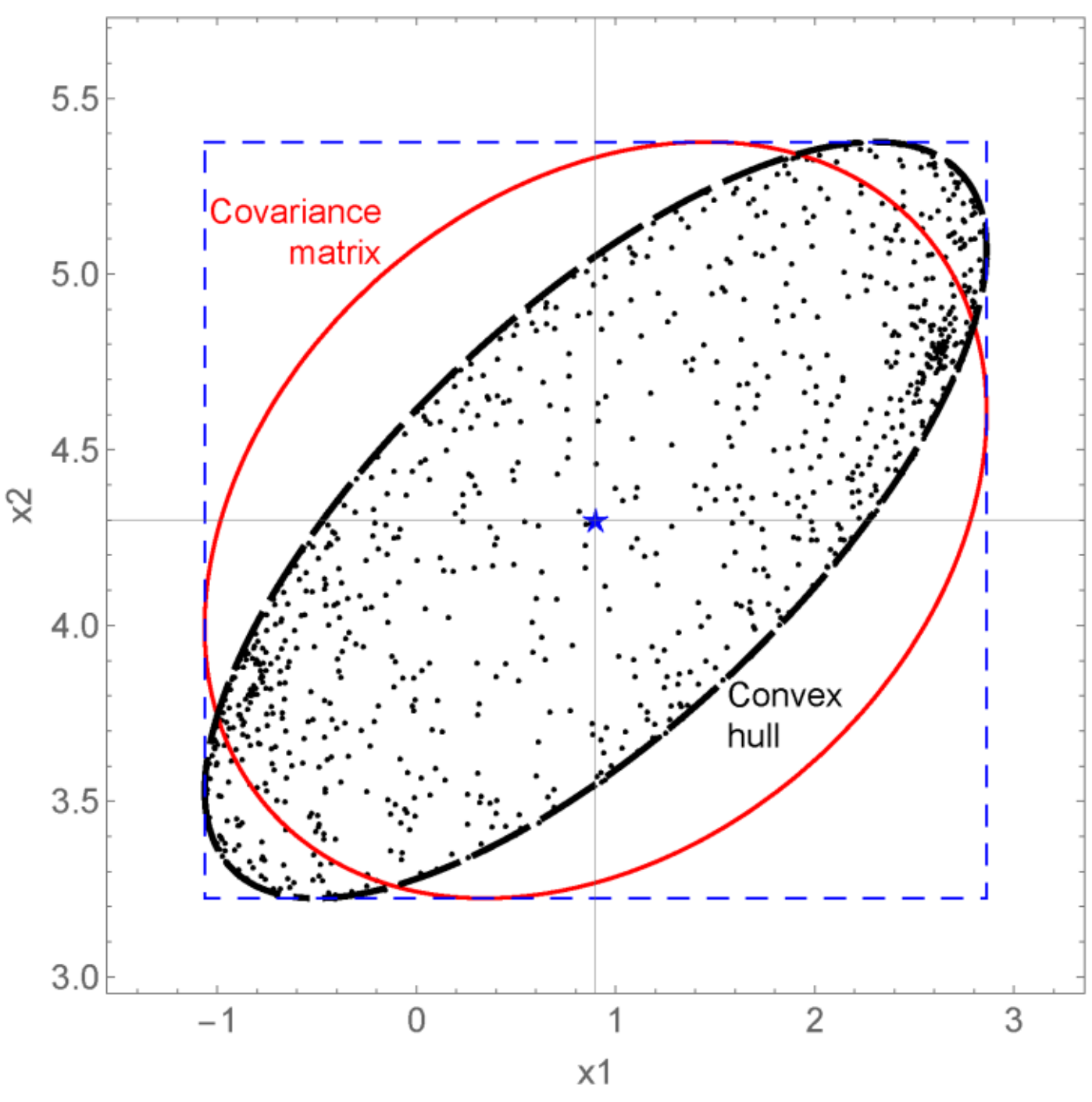

*Figure 3. Reconstructed projections of the 3-dimensional ellipsoid shown in Fig.1.*

Depending on the context, either the Convex Hull (CM) method or Covariance Matrix (CM) method may be preferable for the ellipsoid reconstruction. In the case of the solid 3-dimensional ellipsoid presented in Figure 1, the CM method under/overestimates the spread of the input data points. In the $x_1x_2$ projection of the ellipsoid in Figure 1 and its other projections, the boundary ellipse predicted based on the covariance matrix (red line) has lower eccentricity than the input data. The CH method (black dashed line), on the other hand, traces well the outer boundary of the ellipsoid. Furthermore, the ellipsoid matrix $A_3^{CH}$ reconstructed using the CH method agrees well with the input ellipsoid matrix $A_3^{input}$ used to generate the data, with the relative differences not exceeding 1.5%:

$$A_3^{input} = \begin{pmatrix} 1.439 & -1.607 & 0.626 \\ -1.607 & 2.685 & -0.631 \\ 0.626 & -0.631 & 0.432 \end{pmatrix};\ A_3^{CH} = \begin{pmatrix} 1.453 & -1.621 & 0.634 \\ -1.621 & 2.704 & -0.639 \\ 0.634 & -0.639 & 0.437 \end{pmatrix}.$$

### B. Cross sections for electroweak boson production at the Large Hadron Collider

Our second example establishes a connection to elementary particle physics, where the ellipsoid reconstruction may be employed in large-scale statistical analyses of experimental data from particle colliders. Parton distribution functions (PDFs) quantify the inner structure of the protons in many theoretical calculations in quantum chromodynamics (6). PDFs are published as effective functions dependent on tens to hundreds of free parameters determined from the global analysis of collider data. Knowledge of the statistical distributions of PDF parameters allowed by the experimental data is essential for quantifying the uncertainty on theoretical predictions. In the situations when the parameter distribution is established by stochastic sampling of the multi-dimensional (sometimes 100-dimensional) parameter space (12) (13), the information contained in the PDFs can be effectively compressed using the principal component analysis [PCA] (7) or an alternative compression method (14) (15). Compression of PDFs simplifies their use and combination (16). The Convex Hull ellipsoid reconstruction is similar in its spirit to the PDF compression based on the PCA, while it also reflects deviations from the normality identified by the other compression methods.

As an example of such an application, consider theoretical uncertainties in predicting probabilities (or cross sections) for production of elementary particles in high-energy physics experiments in proton-proton collisions at the Large Hadron Collider (LHC). Rates for production of electroweak bosons $W^{\pm}$, $Z^{0}$, and $H^{0}$ or other heavy particles depend on distributions of partons (quarks and gluons) inside the proton, which are not fully known, but parameterized based on experimental measurements within some uncertainty. If the parton distributions are similar in production of particles A and B, the measurement of the cross section for production of A can constrain the parton distributions in production of B.

We can estimate the probability that the measurement of A will constrain B by plotting pairs of cross sections for A and B for an ensemble of parton distributions. Such plots for production of electroweak bosons at the Large Hadron Collider at beam energy 8 TeV were obtained using Neural Network PDF (NNPDF2.1) parton distributions (17) in Figure 4.

The NNPDF2.1 set provides 1000 forms of PDFs whose parameters are distributed according to the probability prescribed by the pre-LHC data. For each NNPDF parameterization, we plot the total cross sections for two types of bosons ($Z^{0}$ vs. $W^{\pm}$, $W^{+}$ vs. $W^{-}$, $H^{0}$ vs. $Z^{0}$, and $H^{0}$ vs. $W^{\pm}$), and hence obtain a set of 1000 discrete points (indicated by black dots) in 2-dimensional planes of the respective cross sections.

Next, we wish to ask if the predictions based on the NNPDF set follow the Gaussian distribution. If they do, the central regions will be elliptical and concentric for all cumulative probabilities, and thus our ellipsoid reconstruction method may accurately quantify the predictions. For each pair of cross sections, we fit the 68% (red) and 90% (green) ellipses using the Convex Hull Method. [The uncertainties of parton distributions are presented often at the 68% or 90% probability levels.] As we see, for all pairs of cross sections, the 68% and 90% intervals can be approximated by ellipses, but the ellipses are not always concentric. This indicates some deviations from the Gaussian approximation. The reason is that the 1000 NNPDF parton distributions are obtained using a Monte-Carlo statistical method that does not rely on the Gaussian approximation (12) (13). The CH method can be used to reveal deviations from the Gaussian statistics.

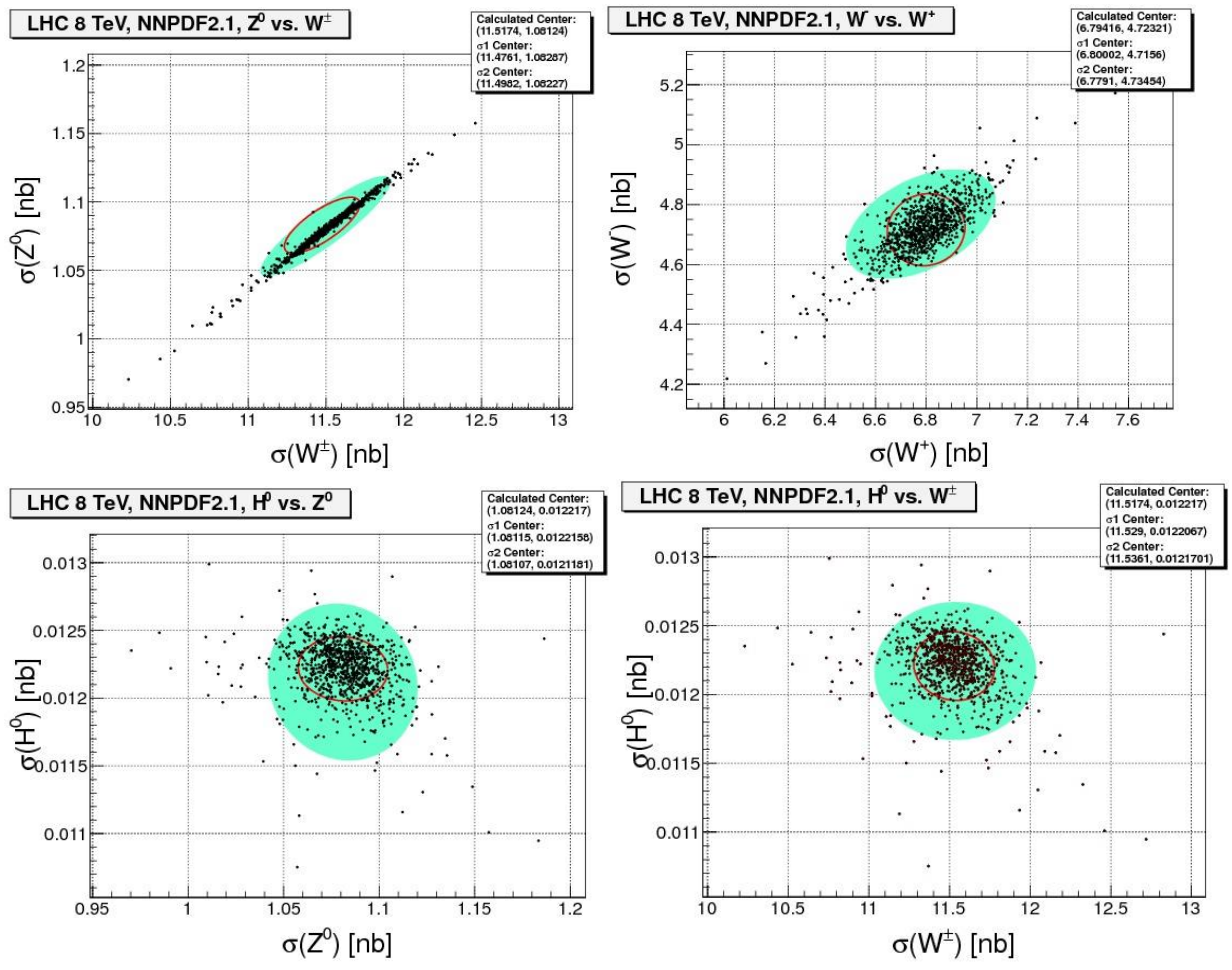

*Figure 4. Next-to-leading order predictions for total cross sections of $W^{\pm}$, $Z^0$, and $H^0$ boson production at the Large Hadron Collider obtained using NNPDF2.1 parameterizations of parton distributions and the Convex Hull Method.*

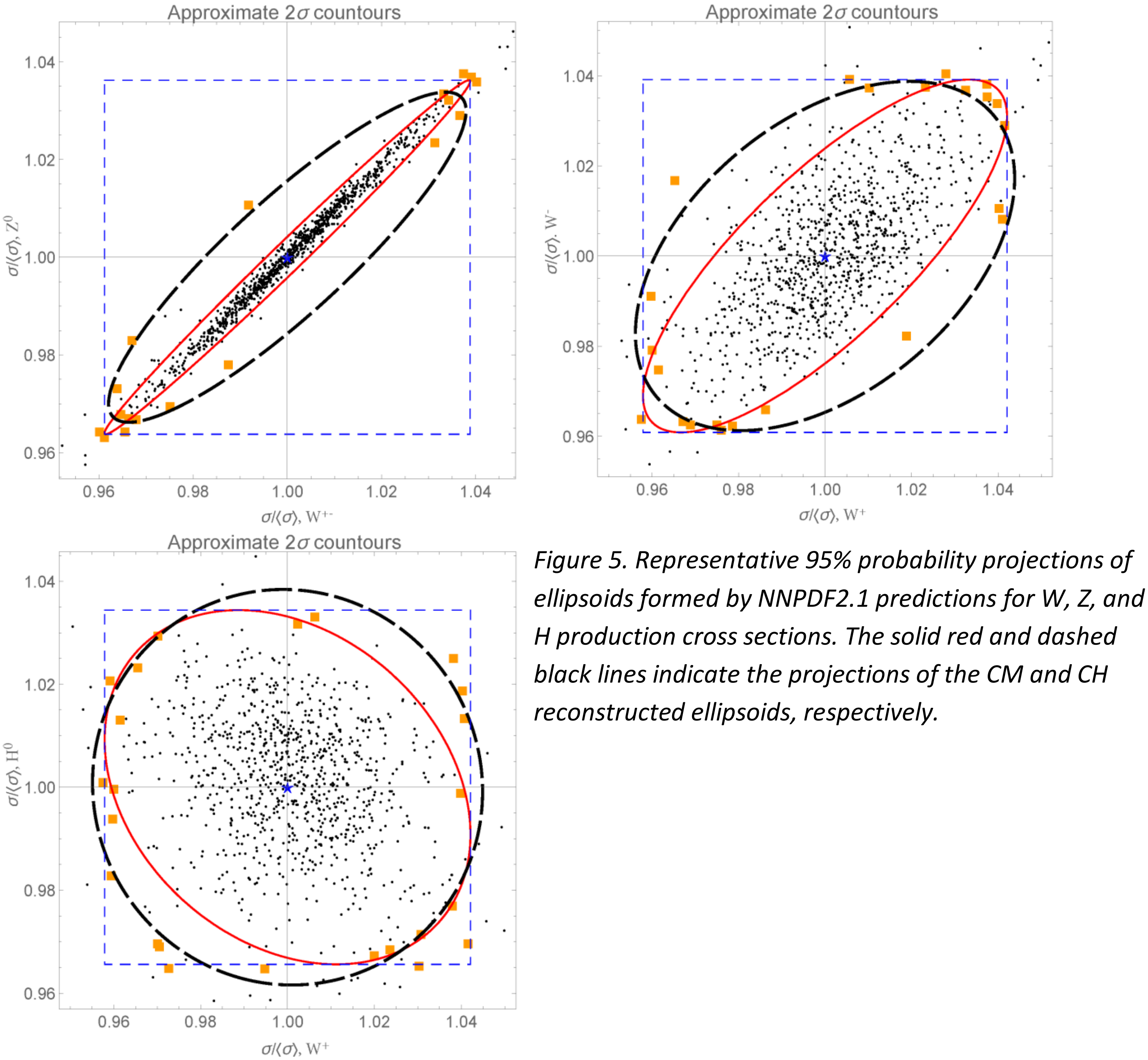

*Figure 5. Representative 95% probability projections of ellipsoids formed by NNPDF2.1 predictions for W, Z, and H production cross sections. The solid red and dashed black lines indicate the projections of the CM and CH reconstructed ellipsoids, respectively.*

The eccentricity of the ellipses quantifies the degree of correlation of the pairs of the cross sections through their PDF dependence (18). Figure 5 shows the correlation ellipses for $W^{\pm} - Z^0$, $W^+ - W^-$, and $W^+ - H^0$ cross sections at the 95% probability level. Here we normalize the cross sections of each type to their mean values over the sample of 1000 replicas, in order to eliminate the dependence on the average magnitude of the production cross sections, which varies depending on the type of the produced particle. We see from the figure that the relative variations due to the parton distributions are of the same order of magnitude for all particle types, not exceeding $\pm 4\%$ in the cross section magnitude at the 95% probability level.

The solid red ellipses in Figure 5 are obtained using the CM method, while the black dashed ellipses are found by fitting the convex hull of the data points enclosed in the overlap of $\pm 2\sigma$ intervals for each cross section of the pair (shown by blue short-dashed lines). Orange squares indicate the points fitted by the convex hull. The 95% CM ellipse automatically lies within the square corresponding to the overlap of the single-variable $\pm 2\sigma$ intervals. The CH ellipse, on the other hand, may go outside of the 95% square. The CH ellipse is more sensitive to outliers and more prone to random fluctuations, especially if it fits only a few points. Rather than fitting only the points exactly on the convex hull, we can fit instead the points within a narrow band around the convex hull in order to suppress the random fluctuations.

Figure 5 shows that the ellipses for $Z^0$ vs. $W^\pm$ and $W^+$ vs. $W^-$ cross sections are very eccentric (highly correlated). A very high correlation normally indicates that the measurement of one cross section will impose tight constraints on the PDFs in the other cross section. The CM method indeed predicts such high correlation. However, we see that a few input data points for these cross sections lie far outside of the CM ellipse. Those on the convex hull are fitted by the CH ellipsoid, but have a small effect on the CM ellipsoid, as the latter is reconstructed from the totality of all points in the Gaussian approximation. Therefore, the deviations from the Gaussian behavior captured by the Convex Hull method result in a *smaller* absolute correlation than according to the Covariance Matrix method.

On the other hand, the cross section for Higgs ($H^0$) boson production is weakly correlated with the $W^\pm$ or $Z^0$ cross sections: measuring $W^\pm$ and $Z^0$ cross sections will not be very helpful for probing the parton distributions relevant for Higgs boson production. The CM and CH methods give comparable predictions for the correlations with the Higgs cross sections.

From 10 independent projections like the ones in Figure 5, we reconstruct the matrices for the 5-dimensional ellipsoid according to Eq. 10. The values of the matrices are

$$A_5^{CM} = 10^4 \times \begin{pmatrix} 61.5 & 42.3 & -96.2 & -8.16 & 0.178 \\ 42.3 & 30.9 & -66.5 & -7.0 & 0.119 \\ -96.2 & -66.5 & 170. & -8.59 & 0.172 \\ -8.16 & -7.0 & -8.59 & 25.4 & -0.411 \\ 0.178 & 0.119 & 0.172 & -0.411 & 0.371 \end{pmatrix},$$

$$A_5^{CH} = 10^4 \times \begin{pmatrix} 0.326 & 0.0509 & -0.165 & -0.136 & 0.0146 \\ 0.0509 & 0.206 & -0.0758 & -0.0639 & 0.00781 \\ -0.165 & -0.0758 & 0.444 & -0.124 & -0.00442 \\ -0.136 & -0.0639 & -0.124 & 0.446 & -0.00434 \\ 0.0146 & 0.00781 & -0.00442 & -0.00434 & 0.248 \end{pmatrix}$$

for the 68% probability level ellipsoids, and

$$A_5^{CM} = 10^4 \times \begin{pmatrix} 15.3 & 10.5 & -24.0 & -2.04 & 0.0446 \\ 10.5 & 7.73 & -16.6 & -1.75 & 0.0298 \\ -24.0 & -16.6 & 42.7 & -2.14 & 0.043 \\ -2.04 & -1.75 & -2.14 & 6.35 & -0.102 \\ 0.0446 & 0.0298 & 0.043 & -0.102 & 0.0929 \end{pmatrix},$$

$$A_5^{CH} = 10^4 \times \begin{pmatrix} 0.228 & 0.0948 & -0.133 & -0.174 & 0.00331 \\ 0.0948 & 0.174 & -0.0954 & -0.146 & 0.00501 \\ -0.133 & -0.0954 & 0.268 & -0.0323 & -0.00409 \\ -0.174 & -0.146 & -0.0323 & 0.386 & -0.00151 \\ 0.00331 & 0.00501 & -0.00409 & -0.00151 & 0.0678 \end{pmatrix}$$

for the 95% probability level ellipsoids.

The diagonal elements $(A_5)_{ii}$ are taken to be the mean values of the $(d-1)=4$ estimates found from independent projections, according to the discussion in Section III. The standard deviations $\delta(A_5)_{ii}$ of these constructed diagonal elements, divided by the mean values $\langle (A_5)_{ii} \rangle$ of the same elements, serve as the estimates of the consistency between the projections. For the matrices above, the ratios $\delta(A_5)_{ii}/\langle (A_5)_{ii} \rangle$ are equal to zero for the CM ellipsoids and range between 0.03 and 0.2 for the CH ellipsoids. The geometric averages of $\delta(A_5)_{ii}/\langle (A_5)_{ii} \rangle$ for the CH ellipsoids are 0.13 (0.1) at the 68% (95%) probability level.

The magnitude of inconsistency of the CH projections may be explained by a small number of points lying on the convex hull. [The 3-dimensional ellipsoid in the previous example contained a large number of points, so its CH projections were practically consistent.] The CH Method selects points on the boundary of the desired two-dimensional probability region. In the projection $x_i x_j$, the selection of $x_i$ points depends on the other dimension $x_j$, as their coordinates must lie both within the probability intervals for $x_i$ and for $x_j$. As the reconstruction algorithm cycles through different projections involving $x_i$, different $x_i$ points will likely be selected, causing some inconsistencies in the coefficients $x_i^2$. Meanwhile, the Covariance Matrix Method does not include a subset-selecting process: all data points are used regardless of the probability level. Thus, in the Covariance Matrix Method, the ellipses are guaranteed to be consistent. In the CH method, the consistency improves by including more points or by fitting the points lying within a band around the convex hull, rather than just on the convex hull itself.

*Table 1. Lengths of the principal semi-axes of the 5-domensional ellipsoids.*

| Probability | Method | Principal semi-axes |
|---|---|---|
| 68% | Covariance Matrix | 0.00063, 0.0017, 0.0093, 0.016, 0.038 |
| | Convex Hull | 0.013, 0.013, 0.020, 0.023, 0.032 |
| 95% | Covariance Matrix | 0.0013, 0.0035, 0.019, 0.033, 0.077 |
| | Convex Hull | 0.013, 0.017, 0.031, 0.038, 0.070 |

Table 1 lists the principal semi-axes of the four reconstructed ellipsoids. In the Covariance Matrix method, the semi-axes of the 95% (2-sigma) ellipsoid are twice as long as the ones for the 68% (1-sigma ellipsoid), as a consequence of the assumed normality of the probability distribution. The lengths span from 0.0013 to 0.077 for the 95% CM ellipsoid, reflecting high eccentricity of the CM ellipsoid in some directions.

The Convex Hull method produces less eccentric ellipsoids because it accounts for the few outlying points that indicate some non-Gaussian behavior. The lengths for the 95% CH ellipsoid range from 0.013 to 0.07, i.e., they are more uniform than the respective lengths of the 95% CM ellipsoid. The ratios of the lengths of the 95% and 68% CH ellipsoids are $0.99, 1.28, 1.55, 1.67$, and 2.16 – very different from 2 for the shortest principal axes.

## V. Conclusion

We presented an algebraic algorithm to obtain a unique, closed solution for the quadratic form of an ellipsoid reconstructed from d-dimensional discrete points using a complete and mutually consistent set of two-dimensional (or, generally, lower-dimensional) orthogonal projections. The reconstruction algorithm requires fitting several two-dimensional ellipses. We explored two approaches to achieving this task: the Convex Hull method, a purely algebraic process that uses cross products and least squares minimization using a generalized eigenvalue equation; and the Covariance Matrix method, which employs strong assumptions of normality to calculate a covariance matrix that determines the ellipse. We then explained how to exploit a simple relationship between their coefficients and those of the inverse of the desired ellipsoid's quadratic form. In outlining this process, we proved that it is guaranteed to lead to a unique solution.

Finally, we realized the implementation of our algorithm in a Mathematica program and applied it to reconstruction of a three-dimensional solid ellipsoidal body as well as to a statistical distribution of cross sections for elementary particle production at the LHC. These applications illustrate when the Convex Hull and Covariance Matrix methods may produce different results. The suitability of each method depends on the context. The Convex Hull method is sensitive to outliers and deviations from the Gaussian behavior, though measures may be taken to suppress this sensitivity to a certain extent. In the non-Gaussian cases, it may give inconsistent coefficients for the ellipsoid's quadratic form. In general, the Convex Hull method estimates correlations between the parameters more conservatively than the Covariance Matrix method, which is less sensitive to outliers, produces perfectly consistent closed forms of elliptical projections, and can provide very aggressive predictions for correlations between parameters.

Each method performs well under a certain set of circumstances, and comparing the ellipsoids determined by both methods serves as a normality test of the underlying probability distribution. In the above example of the electroweak particle production at the LHC , the Convex Hull method indicates a weaker correlation between the production cross sections of $W^{\pm}$ and $Z^0$ bosons than would be estimated by the commonly used Covariance Matrix formalism. The difference arises because of the non-Gaussian effects revealed by the NN parton distributions and may have practical implications for constraining precision measurements of $W^{\pm}$ bosons by the "benchmark" measurements of $Z^0$ bosons.

As the basis of this algorithm is purely mathematical, it can be applied in many fields of science. The development of a program that fits ellipsoids to sets of discrete multi-dimensional data has proved to be a useful way of determining correlations between parton distributions and particle production. This is just one application of the algorithm discussed; countless more exist. The research's goal of producing a program that can efficiently fit ellipsoids to sets of discrete multi-dimensional data was accomplished, as the coded implementation of the algorithm has been tested and proven to be accurate.

### Acknowledgments

This work was supported in part by the SMU Senior Engaged Learning Fellowship, undergraduate research assistantships, and by the U.S. Department of Energy under Grant No. DE-SC0010129.